\documentclass[journal = jctcce,manuscript=article]{achemso}
\usepackage[utf8]{inputenc}
\usepackage[T1]{fontenc} 
\usepackage{times}
\usepackage[none]{hyphenat} 
\usepackage{array} 
\usepackage{multirow} 
\DeclareUnicodeCharacter{0301}{\'{e}}
\usepackage{amsmath}

\usepackage{threeparttable}
\usepackage{graphicx} 
\usepackage{caption}
\usepackage[font=large]{subfig}

\usepackage{hyperref}
\usepackage{amssymb}
\usepackage{textcomp}
\usepackage{gensymb}
\usepackage{xcolor}

\makeatletter
\newcommand*{\addFileDependency}[1]{
  \typeout{(#1)}
  \@addtofilelist{#1}
  \IfFileExists{#1}{}{\typeout{No file #1.}}
}

\makeatother

\title{Externally corrected CCSD with renormalized perturbative triples (R-ecCCSD(T)) and density matrix renormalization group and selected configuration interaction external sources}
\author{Seunghoon Lee}
\affiliation{Division of Chemistry and Chemical Engineering, California Institute of Technology, Pasadena, California 91125, USA}
\email{slee89@caltech.edu}

\author{Huanchen Zhai}
\affiliation{Division of Chemistry and Chemical Engineering, California Institute of Technology, Pasadena, California 91125, USA}

\author{Sandeep Sharma}
\affiliation{Department of Chemistry, The University of Colorado at Boulder, Boulder, Colorado 80302, USA}
\email{sandeep.sharma@colorado.edu}

\author{C. J. Umrigar}
\affiliation{Laboratory of Atomic and Solid State Physics, Cornell University, Ithaca, New York 14853, USA}
\email{cyrusumrigar@gmail.com}

\author{Garnet Kin-Lic Chan}
\affiliation{Division of Chemistry and Chemical Engineering, California Institute of Technology, Pasadena, California 91125, USA}
\email{gkc1000@gmail.com}

\begin{document}

\begin{abstract}
We investigate the renormalized perturbative triples correction together with the externally corrected coupled-cluster singles and doubles (ecCCSD) method. 
We take the density matrix renormalization group (DMRG) and heatbath CI (HCI) as external sources for the ecCCSD equations. 
The accuracy is assessed for the potential energy surfaces of H${}_2$O, N${}_2$, and F${}_2$.
We find that the triples correction significantly improves on ecCCSD and we do not see any instability of the renormalized triples with respect to dissociation. 
We explore how to balance the cost of computing the external source amplitudes with respect to the accuracy of the subsequent CC calculation. 
In this context, we find that very approximate wavefunctions (and their large amplitudes) serve as an efficient and accurate external source.
Finally, we characterize the domain of correlation treatable using the externally corrected method and renormalized triples combination studied in this work via a well-known wavefunction diagnostic.
\end{abstract}
 
\section{Introduction}

Electronic structure methods that can efficiently handle both static and dynamic correlation remain an important area of investigation.
Because there is a wide spectrum of strongly correlated problems, ranging from mildly "quasi"-degenerate scenarios (e.g. in the electronic structure of diradicals \cite{abe2013diradical, nakano2017diradical}) to extensively near-degenerate problems (e.g. in the electronic structure of multi-centre transition metal clusters \cite{kurashige2013entangled, sandeep2014fe2s2, zhendong2019pcluster})  a variety of theoretical strategies have been proposed.



For highly-degenerate problems, it is common to combine dynamic correlation methods with an explicit treatment of a multi-reference state.
The representation of the multi-reference state can range from an exact complete active space (CAS) representation~\cite{das1972new, werner1980quadratically, werner1985second} (for small numbers of orbitals), to density matrix renormalization group (DMRG)~\cite{white1992real, white1992density, white1993density, white1999ab, chan2002highly, legeza2003controlling, chan2004algorithm, legeza2008applications, chan2009density, marti2010density, chan2011density, sharma2012spin, wouters2014density, szalay2015tensor, yanai2015density} (e.g. when almost all orbitals are degenerate), to selected configuration interaction~\cite{ivanic2001identification, huron1973iterative, GarSceGinCafLoo-JCP-18,LooSceBloGarCafJac-JCTC-18,GarSceLooCaf-JCP-17, buenker1974individualized, evangelisti1983convergence, harrison1991approximating, STEINER1994263, Neese2003, ABRAMS2005121, BYTAUTAS200964, evangelista2014driven, knowles2015compressive, schriber2016communication, tubman2016deterministic, liu2016ici, caffarel2016using, Holmes_2016, sandeep2017, LiOttHolShaUmr-JCP-18} and Monte Carlo approximations~\cite{booth2009fermion, cleland2010communications, ghanem2019unbiasing, sabzevari2018improved, neuscamman2016subtractive} for intermediate cases. 
On top of these, various flavours of perturbation theory~\cite{andersson1990second, angeli2001introduction, angeli2001n, angeli2002n, neuscamman2010review, yanai2010multireference, kurashige2011second, sharma2014communication, sharma2015multireference, guo2016n, sokolov2017time, sharma2017semistochastic, guo2018communication}, configuration interaction~\cite{lischka1981new, szalay2012multiconfiguration, lischka2001high, saitow2015fully}, and
exponential approximations~\cite{paldus1999adv, kong2009state, jeziorski2010multireference, sharma2015multireference} have been explored.
However, the combination of dynamic correlation
with multi-reference representations is not straightforward and usually leads to added conceptual, implementation, and computational complexity.

For quasi-degenerate problems, an alternative strategy can be used, which  incorporates a limited amount of static correlation on top of an existing SR  method. This has been particularly popular
in conjunction with SR coupled cluster methods.~\cite{vcivzek1966correlation, vcivzek1969use, vcivzek1971correlation, paldus1972correlation, bartlett1995modern, shavitt2009many} Some examples include 
variants of tailored coupled cluster~\cite{kinoshita2005coupled, hino2006tailored, lyakh2011tailored, veis2016coupled, faulstich2019numerical, vitale2020fciqmc} and externally corrected coupled cluster methods.~\cite{paldus1994valence, planelles1994valence, planelles1994valence2, paldus2017externally} The computational cost of such SR static correlation methods is often lower than that of true MR dynamic correlation methods and thus large active spaces become affordable. However, the approximations are limited to problems with only a modest amount of degeneracy. 

In this work, we will focus on quasi-degenerate problems, and in particular, we will investigate the externally corrected coupled cluster method.~\cite{paldus1994valence, planelles1994valence, planelles1994valence2, paldus2017externally}
This extracts  static correlation from a MR method by using the MR wavefunction as an "external" source of higher order coupled cluster amplitudes. For example, in the ecCCSD approximation, the $T_3$ and $T_4$ amplitudes are extracted from the external source and a new set of $T_1$ and $T_2$ amplitudes are computed in their presence. Should the $T_3$ and $T_4$ amplitudes be exact, then the $T_1$ and $T_2$ amplitudes  {and the energy} will be exact. A different, spiritually related, approximation is tailored CCSD~\cite{kinoshita2005coupled, hino2006tailored, lyakh2011tailored}, which has been of renewed interest of late~\cite{veis2016coupled, faulstich2019numerical, vitale2020fciqmc}. Here, instead of higher order cluster amplitudes, the large (active space) $T_1$ and $T_2$ amplitudes are fixed from an external source. 

The ecCCSD method has a long history and external sources, ranging from unrestricted Hartree-Fock~\cite{paldus1982cluster, paldus1983bond, paldus1984approximate, piecuch1996approximate} to CASSCF and CASCI~\cite{peris1997single, li1997externally, peris1998externally, stolarczyk1994complete, xu2015externally} and, most recently, full configuration interaction quantum Monte Carlo\cite{deustua2018}, have been used. One of the more successful applications 
of ecCCSD is the reduced multireference (RMR) CCSD method\cite{li1997, li2000truncated}, which uses a MRCISD wave function as the external source.
 RMR CCSD(T), which incorporates some of the residual dynamic correlation through perturbative triples, has also been studied.\cite{li2006}
Despite the promising performance of RMRCCSD(T) in several studies~\cite{li1998dissociation, li1998singlet, li1999simultaneous, li2000reduced, li2000effect, li2006truncated, li2006singlet, li2007reduced, li2008coupled}, it suffers from two main limitations. First, conventional MRCISD can only be applied for modest sizes of active spaces (typically, up to about $16$ orbitals as limited by the exact CAS treatment). Second, the (T) correction, although not divergent like  its single-reference counterpart, still overcorrects the dynamical correlation in the bond-stretched region.\cite{li2008n2}

In this work, we make two modifications to ecCCSD to overcome and ameliorate the above limitations.
First, we utilize variational DMRG and HCI wave functions as external sources for ecCCSD.
This allows for the use of larger quasidegenerate active spaces, of the size typically treated by DMRG and HCI.
Second, we explore the renormalized perturbative triples correction. This has been shown in the single reference setting to ameliorate the overcorrection of standard perturbative triples \cite{kowalski2000mmcc}, without affecting the computational scaling.

We describe the use of DMRG and HCI as external sources for ecCCSD  in Section~\ref{sec:ecCCSD}. 
It is possible to create a near-exact method by using a near-exact external source.
Since near-exact external sources can be computed by DMRG and HCI for the small systems we employ as
test cases in this paper,
the critical question is not simply the accuracy of the method, but the balance between cost and accuracy.
To this end, we explore a variety of approximate treatments of the external DMRG and HCI sources as discussed in Sec.~\ref{sec:ea}.
The various triples approximations for  ecCCSD are discussed in Section \ref{sec:recccsdt}.
Computational details are provided in Section \ref{sec:compdetail} and the accuracy of the renormalized perturbative triples correction is assessed for
 three potential energy surfaces (PESs) in Sections \ref{sec:fm}-\ref{sec:fpal}. We characterize the range of quasi-degenerate correlations captured in this work in Section \ref{sec:diag}. Finally we discuss the limitations of this method in Section \ref{sec:limit} and summarize our findings in Section \ref{sec:conclusions}.

\section{Theory}

In this work, we use a reference configuration $| \Phi \rangle$ with the same occupancy as the 
Hartree-Fock (HF) determinant.
It is useful to define a projection operator onto the space of $k$-tuply excited configurations relative to the reference; we denote this $Q_k$. The external source is used to provide an important subset of the triply and quadruply excited configurations; the projector onto this subset is denoted $Q_k^{ec}$, and its complement is $Q_k^c$.
Thus 
\begin{alignat}{2}
    Q_k & = Q^{\rm ec}_k + Q^{\rm c}_k, \ \ & k&=3,4, \label{eq:p34} 
\end{alignat}

\subsection{Externally corrected CCSD with DMRG and HCI wave functions} \label{sec:ecCCSD}
In ecCCSD, the coupled cluster operator $T$ is given by
\begin{align}
    T = T_1 + T_2 + Q^{\rm ec}_3 T_3 + Q^{\rm ec}_4 T_4. \label{eq:t}
\end{align}
Here, $T_n, \ n=1,2,3,4, $ are the $n$-fold cluster operators.
In this work, we extract $Q^{\rm ec}_k T_k$, $k=3,4$ from the DMRG or the HCI variational wave function.
For the DMRG wave function, the triply and quadruply excited configurations $D_p$ that define $Q^{\rm ec}_k$, $k=3, 4$ are chosen
to be those where the magnitude of the CI coefficient $c_p$  is above a threshold, i.e.
\begin{align}
    |c_p| > s \sqrt{\omega}, \label{eq:dmrg_V}
\end{align}
where $s$ is an arbitrary scaling factor and $\omega$ is the largest discarded weight of the density matrix at the maximum bond dimension in two-dot DMRG sweeps (carried out without noise). An efficient algorithm to convert a matrix-product state to CI coefficients above a given threshold is described in Appendix \ref{apdx:mpsci}. For the HCI variational wave function, $D_p$ is included in the projectors $Q^{\rm ec}_k$, $k=3, 4$ using the heatbath algorithm with a threshold $\epsilon$,\cite{Holmes_2016} i.e., it is included if
\begin{align}
    | \langle D_p | H | D_q \rangle c_q | > \epsilon, \label{eq:shci_V}
\end{align}
for at least one determinant $D_q$ which is already in the variational space. The extracted CI coefficients are then converted into cluster amplitudes.

The $T_1$ and $T_2$ amplitudes are obtained by solving the ecCCSD  equations using fixed $Q^{\rm ec}_3T_3$ and $Q^{\rm ec}_4T_4$,
\begin{equation}
        0 = (Q_1 + Q_2) (H_N e^{T_1 + T_2 + Q^{\rm ec}_3 T_3 + Q^{\rm ec}_4 T_4})_C | \Phi \rangle,
\end{equation}
where $H_N$ is the Hamiltonian in  normal-ordered form, and the subscript $C$ denotes the connected part of the corresponding operator expression. With the relaxed $T_1$ and $T_2$, the ecCCSD correlation energy of the ground state is obtained as
\begin{align}
    \Delta E^{\rm ecCCSD}_0 = \langle \Phi \vert (H_N e^{T_1 + T_2 + Q^{\rm ec}_3 T_3 + Q^{\rm ec}_4 T_4})_C \vert \Phi \rangle.
\end{align}

\subsection{Approximations in the external  source} \label{sec:ea}
While DMRG and HCI can, in small molecules, be a source of nearly exact $T_3$ and $T_4$ amplitudes even in the full
orbital space,
this provides no computational advantage as such calculations are more expensive than the subsequent coupled cluster calculation.
Consequently, it is important to balance the cost of the external source calculation and that of the subsequent coupled cluster calculation by making approximations in the external source. This introduces the additional complication that one must ensure that errors introduced into the external source do not lead to unacceptable errors in the final coupled cluster calculation. 

In this work, we consider {six} different types of approximate external sources with different sizes of active spaces and different values of parameters summarized in Table \ref{table:apprx_ext}.
Type I uses CASSCF-like external sources in minimal active spaces. Since the minimal active space of F$_2$, i.e., two electrons in two orbitals (2e,2o), does not contain $T_3$ and $T_4$, we perform minimal active space ecCC calculations for only H$_2$O and N$_2$ (with (4e,4o) and (6e,6o), respectively). 
These provide amplitudes that are very close to the exact CASSCF amplitudes.
However, these amplitudes lack the relaxation that comes from allowing excitations within a larger space
of orbitals, and of course the amplitudes outside the minimal active space are completely absent.
Type III uses CASCI-like external sources (the orbitals are not optimized to save computer time)
in larger active spaces ((8e,18o), (10e,16o), and (14e,16o) for H$_2$O, N$_2$, and F$_2$, respectively).

In either case, one can potentially introduce bad external amplitudes if the effect of relaxation on the amplitude
upon going to a larger space is large relative to the size of the amplitude (e.g. changes its sign).
Thus, we study also Type II and Type IV external sources which are similar to Type I and Type II external sources respectively
except that they employ an additional threshold to screen out all except the largest $T_3$ and $T_4$ amplitudes. 
The absolute values of the $T_3$ and $T_4$ elements {at the most stretched geometry of each molecule} are sorted in a single large vector. The norm of the vector is computed. Only the largest elements of the vector are retained such that
the resulting norm is more than 80\% of the norm of the full vector. 
{Along PESs of each molecule, we used the same set of elements of $T_3$ and $T_4$ (but with the appropriate values for each geometry) as the external sources, to maintain the smoothness of the PESs.}


{As discussed in Section~\ref{sec:fpal}, the type-III and type-IV sources improve upon the PESs obtained from the type-I and type-II sources,
but the DMRG calculations to obtain the sources} incur a higher computational cost than the subsequent CC calculations.
To reduce the cost, we have also tried type-V sources, which employ loosely converged DMRG wave functions with small bond dimensions of $M=25,50,$ and $100$.
Finally, type-VI sources employ large thresholds $\epsilon=0.01,$ and $0.003$ to obtain loosely converged HCI wave functions in the full orbital spaces.
This combination has the advantage that it can be considered a black-box method wherein a single parameter $\epsilon$ controls the tradeoff between accuracy and cost.

\begin{table}[htp]
\caption{Six types of approximate external sources. Detailed explanation to be found in the main text. }
\begin{tabular}{cccll}\hline \hline
\multirow{2}{*}{Type}& \multirow{2}{*}{Method}
                            & \multirow{2}{*}{Active space} & \multicolumn{2}{c}{Cut-off parameters}            \\
                    &       &                               & \multicolumn{1}{c}{Wave function} & \multicolumn{1}{c}{Cluster amplitude}             \\ \hline
\multirow{2}{*}{I}  &  DMRG &  \multirow{2}{*}{Minimal${}^a$}     & $M=2000$          & $100$\% with $s=0.1$  \\
                    &  HCI  &                                     & $\epsilon=10^{-7}$& $100$\%               \\ \\ 
\multirow{2}{*}{II} &  DMRG &  \multirow{2}{*}{Minimal${}^a$}     & $M=2000$          & $80$\% with $s=0.1$   \\
                    &  HCI  &                                     & $\epsilon=10^{-7}$& $80$\%                \\ \\ 
               III  &  DMRG &  Larger${}^b$      & $M=2000$          & $100$\% with $s=0.1$  \\ \\ 
                IV  &  DMRG &  Larger${}^b$      & $M=2000$          & $80$\% with $s=0.1$   \\ \\ 
\multirow{2}{*}{V}  &  \multirow{2}{*}{DMRG}& \multirow{2}{*}{Larger${}^b$} & \multirow{2}{*}{$M=25,50,100$} & $80$\% \ \ for H${}_2$O and N${}_2$, \\
                    &                       &                               &                                & $100$\% for F${}_2$ with $s=0.01$  \\ \\
  VI  &  HCI& Full${}^c$ & $\epsilon=0.01, 0.003$ & $100$\%  \\ \hline \hline
\end{tabular} \label{table:apprx_ext}
\begin{alignat*}{6}
    & {}^{a} \mathrm{(4e,4o)}  \ & \mathrm {for} \ \mathrm{H}{}_2{\mathrm O}, \ \ & \mathrm{(6e,6o)  } \ & \mathrm {for} \ \mathrm{N}{}_2  \ \ & & \nonumber \\ 
    & {}^{b} \mathrm{(8e,18o)} \ & \mathrm {for} \ \mathrm{H}{}_2{\mathrm O}, \ \ & \mathrm{(10e,16o)} \ & \mathrm {for} \ \mathrm{N}{}_2, \ \ & \mathrm {(14e,16o)}\ & \mathrm {for} \ \ \mathrm{F}{}_2 \nonumber \\
    & {}^{c} \mathrm{(10e,58o)}\ & \mathrm {for} \ \mathrm{H}{}_2{\mathrm O}, \ \ & \mathrm{(14e,60o)} \ & \mathrm {for} \ \mathrm{N}{}_2, \ \ & \mathrm {(14e,58o)}\ & \mathrm {for} \ \ \mathrm{F}{}_2 \nonumber
\end{alignat*}
\end{table}

\subsection{Perturbative triples corrections} \label{sec:recccsdt}
Expressions for the standard, renormalized, and completely renormalized  perturbative triples corrections can be written down in analogy with
their single reference definitions~\cite{kowalski2000mmcc}.
We first define the completely renormalized (CR)-ecCCSD(T) correction.
We use the state $|\Psi^{\rm ecCCSD(T)} \rangle$ defined as 
\begin{align}
    \vert \Psi^{\rm ecCCSD(T)} \rangle & = ( 1 + T_1 + T_2 + Q^{\rm ec}_3 T_3 + Q^{\rm c}_3 T^{[2]}_3 + Q^{\rm c}_3 Z_3 ) \vert \Phi \rangle, \\
    T^{[2]}_3 \vert \Phi \rangle & = R^{(3)}_0 (V_N T_2)_C \vert \Phi \rangle, \\
    Z_3 \vert \Phi \rangle & = R^{(3)}_0 V_N T_1 \vert \Phi \rangle.
\end{align}
where $V_N$ is the two-body part of the Hamiltonian in  normal-ordered form and $R^{(3)}_0$ denotes the three-body component of the reduced resolvent operator in  many-body perturbation theory, given by  differences of orbital energies in the denominator.\cite{shavitt2009many}
The resulting formula for the CR-ecCCSD(T) energy correction is
\begin{align}
    \delta_0^{\rm CR-ecCCSD(T)} & = \frac{ \langle \Psi^{\rm ecCCSD(T)} | Q^{\rm c}_3 ( H_N e^{T} )_C | \Phi \rangle 
    }{ \langle \Psi^{\rm ecCCSD(T)} | e^{T} | \Phi \rangle }.
\end{align}

The energy corrections for renormalized (R)-ecCCSD(T) and ecCCSD(T) can be obtained by taking 
the lowest-order estimates of the correction and by assuming the denominator to be one, i.e., 
\begin{align}
    \delta_0^{\rm R-ecCCSD(T)} & = \frac{ \langle \Psi^{\rm ecCCSD(T)} | Q^{\rm c}_3 (V_N T_2)_C | \Phi \rangle }{ \langle \Psi^{\rm ecCCSD(T)} | e^{T} | \Phi \rangle }, \label{eq:Recc} \\
    \delta_0^{\rm ecCCSD(T)} & = \langle \Psi^{\rm ecCCSD(T)} | Q^{\rm c}_3 (V_N T_2)_C | \Phi \rangle.
\end{align}

Unlike perturbative triples without external correction (as in CCSD(T)), the approximate perturbative expression $T_3^{[2]}$ is only evaluated for determinants omitted in the external source. Thus it is not expected to diverge as long as the external source includes all degeneracies. Nonetheless, it can still overestimate the triples correlation. The role of the denominator in the "renormalized" triples approximations is to rescale this correction, which can be expected to
reduce the overestimation.

\section{Computational Details} \label{sec:compdetail}

All CC calculations were performed using cc-pVTZ basis sets.~\cite{dunning1989a}
The ecCC calculations were performed with the six different types of external sources summarized in Table~\ref{table:apprx_ext} and discussed in Section~\ref{sec:ea}. {The CASSCF-like external sources (type I and II) used natural orbitals in the minimal active spaces and CASSCF orbitals for the core and external spaces, while
the CASCI-like external sources (type III-VI) used HF orbitals.
}

All CC calculations were carried out using a local version of PySCF \cite{PYSCF} interfaced with StackBLOCK \cite{garnet2002, garnet2004, ghosh2008, sandeep2012} for DMRG and Arrow \cite{LiOttHolShaUmr-JCP-18, ShaHolJeaAlaUmr-JCTC-17, HolTubUmr-JCTC-16} for HCI.
We used Dice \cite{ShaHolJeaAlaUmr-JCTC-17, HolTubUmr-JCTC-16, smith2017} to get accurate SHCI PESs.

\section{Numerical results}

\subsection{PESs with type-I external sources} \label{sec:fm}
\begin{figure}[htp]
\centering
\caption{PESs of H${}_2$O and N${}_2$ in the top and bottom panels, respectively, obtained with the CC methods and the ecCC methods using the type-I external sources. The type-I sources correspond to near exact wave functions from the minimal active space, and use all the amplitudes of $T_3$ and $T_4$. The blue, green, and red solid lines are the ecCCSD, ecCCSD(T), and R-ecCCSD(T) PESs, respectively. These are to be compared with the PESs of CC represented as dotted lines. The black lines are accurate PESs obtained by SHCI.
}
\includegraphics[width=0.7 \textwidth,height=20cm,keepaspectratio]{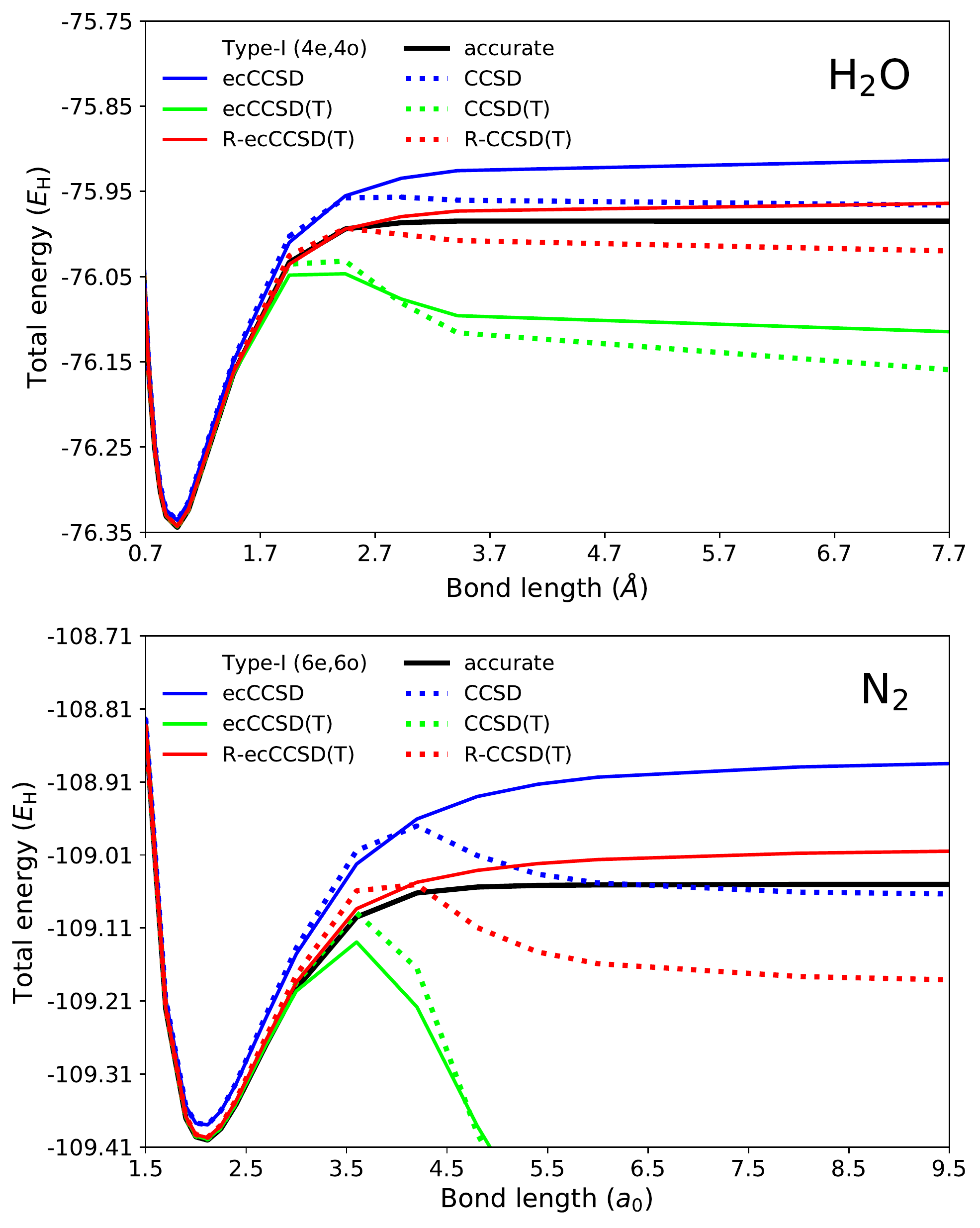} \label{fig:fm}
\end{figure}

The dissociation PESs of H${}_2$O and N${}_2$, shown in Fig.~\ref{fig:fm}, were obtained by the ecCC methods using the type-I external sources in Table~\ref{table:apprx_ext} (i.e. near exact wave functions in the minimal active spaces and all amplitudes of $T_3$ and $T_4$).
As expected, the ecCC curves with tightly converged HCI and DMRG external sources are almost identical. Thus, we show the PESs of ecCC using only one of these external sources  (colored solid lines) and the PESs of CC (colored dotted lines) in Fig. \ref{fig:fm}. These are compared against  accurate PESs represented as black lines obtained by SHCI in the full space. The accurate energies from SHCI are given in the Supporting Information. The mean absolute errors (MAE) and the non-parallelity errors (NPE) of the PESs are listed in Table \ref{table:a}.


The CCSD curves (blue dotted lines) have an unphysical {dip due to an inadequate treatment of static correlation.}
This problem is completely eliminated within ecCCSD (blue  solid line). However, there are significant errors with respect to the black curve at large distances, giving a (MAE,NPE) of ($24.0, 65.9$) m$E_{\rm H}$ for H${}_2$O and ($64.1, 149.9$) m$E_{\rm H}$ for N${}_2$.

We see that the (T) correction captures much of the missing dynamic correlation, such that the green ecCCSD(T) curves reach an accuracy of 1 mE${}_H$  around the equilibrium geometry. However, when the bond is stretched, the (T) correction overestimates the dynamic correlation, leading to another unphysical dip in the PESs. 
Although this overestimation can easily be reduced by increasing the size of the active space for the
small systems treated here, this would be very expensive for large systems.
Alternatively, we can use the renormalized triples formula to damp the (T) correction in Eq. (\ref{eq:Recc}).
R-ecCCSD(T) (the solid red curves) completely removes the unphysical dips in the PESs.
These attain (MAE,NPE) = ($3.9, 23.1$) and ($12.8, 41.4$) m$E_{\rm H}$ for H${}_2$O and N${}_2$, respectively.

\subsection{PESs with type-II external sources} \label{sec:pm}
\begin{figure}[htp]
\centering
\caption{PESs of N${}_2$ obtained by the ecCC methods using all the external amplitudes (type-I) and some of the largest external amplitudes (type-II) from the minimal active space (solid and dotted lines, respectively). Other descriptions are the same as in Fig.\ref{fig:fm}}
\includegraphics[width=0.7\textwidth,height=20cm,keepaspectratio]{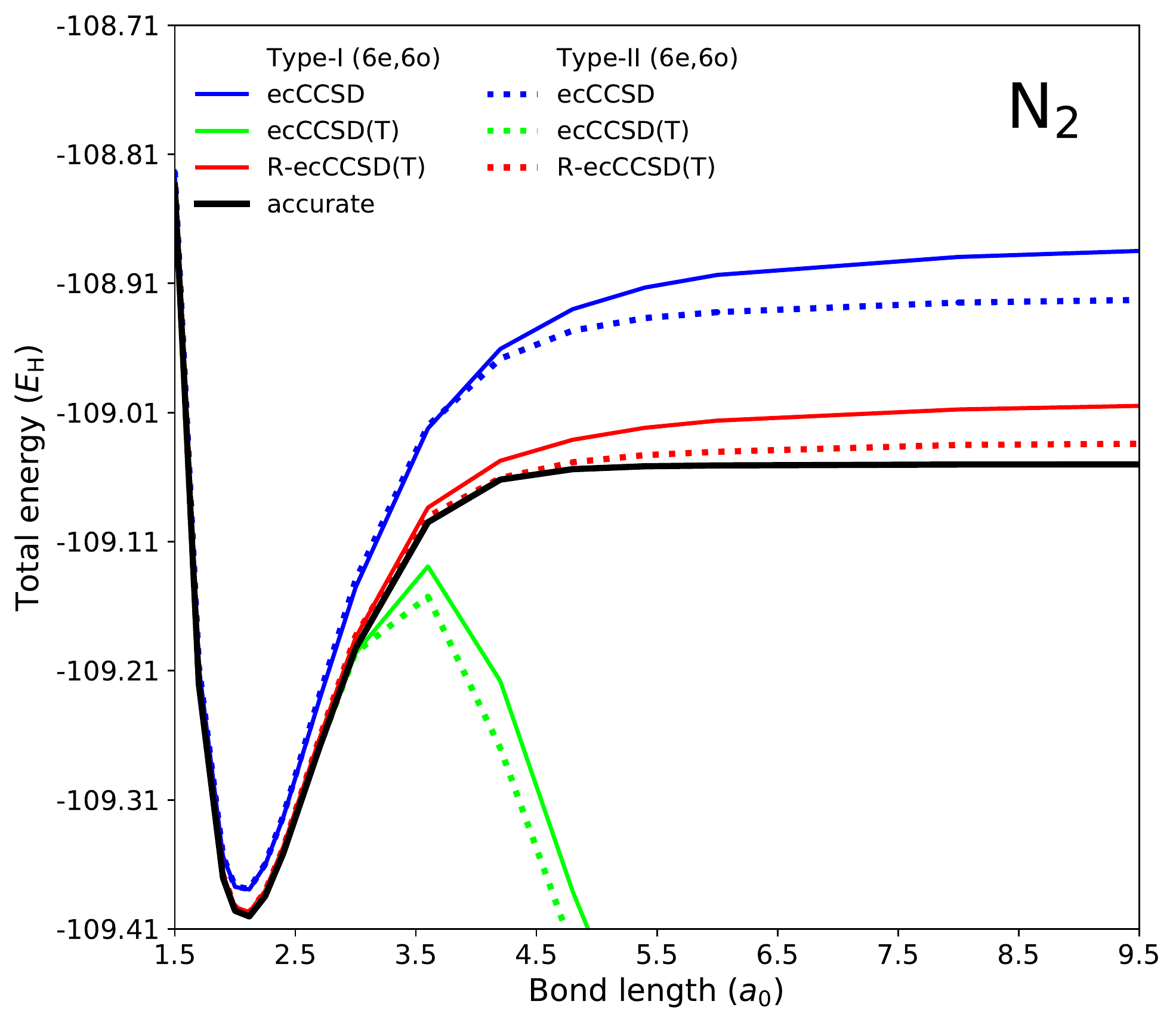} \label{fig:pm}
\end{figure}

The difference between the type-I and type-II external sources is that the former use all the amplitudes of $T_3$ and $T_4$ while the latter use only a small number of the largest amplitudes, which contribute about 80\% of the total weight of $T_3$ and $T_4$ (see Table~\ref{table:apprx_ext} and Section~\ref{sec:ea}).
For the minimal active space of H${}_2$O, i.e. (4e,4o), the external source contains only one non-zero  $T_4$ amplitude, and all elements of $T_3$ are zero to within numerical noise. 
Thus, the PESs of ecCC using 80\% of the external amplitudes (type-II) are almost identical to those using 100\% of the external amplitudes (type-I).
On the other hand, for the minimal active space of N${}_2$, i.e. (6e,6o), the external source contains several
large $T_3$ and $T_4$ amplitudes.
At the stretched geometry corresponding to a bond length of $10 a_0$, 80\% of the total $T_3$ and $T_4$ amplitude weight is recovered by the nine largest elements of $T_4$.
Figure \ref{fig:pm} shows that the type-II external sources, although using fewer amplitudes, improve the PESs of ecCCSD and R-ecCCSD(T) relative to the type-I sources. The PES of R-ecCCSD(T) displays a (MAE,NPE) = ($5.4, 13.9$) m$E_{\rm H}$.

One concern with  partial use of the amplitudes in the minimal active space is the possibility of divergence in (T). 
Assuming a HF occupation of the natural orbitals, the recomputed orbital energies (diagonal parts of the Fock matrix) of the holes and those of the particles are such that the system retains a sizable gap.  N${}_2$ at the $10.0a_0$ bond length, for example, has a gap of around $0.1$ Hartrees between the
three occupied orbitals and the three unoccupied orbitals.
This prevents the divergence of the (T) and the renormalized (T) corrections at the bond lengths we considered (although (T) largely overestimates dynamic correlation).

\subsection{PESs with type-III and type-IV external sources} \label{sec:fpa}
\begin{figure}[htp]
\centering
\caption{PESs of H${}_2$O, N${}_2$, and F${}_2$ in the top, middle, and bottom panels, respectively, for the ecCC methods using the larger than minimal active spaces and the near exact external sources. The solid and dotted lines correspond to PESs obtained using all external amplitudes of $T_3$ and $T_4$ (type-III) and only the largest amplitudes of $T_4$ (type-IV), respectively.
The descriptions are otherwise the same as those in Fig.\ref{fig:pm}.}
\includegraphics[width=\textwidth,height=20cm,keepaspectratio]{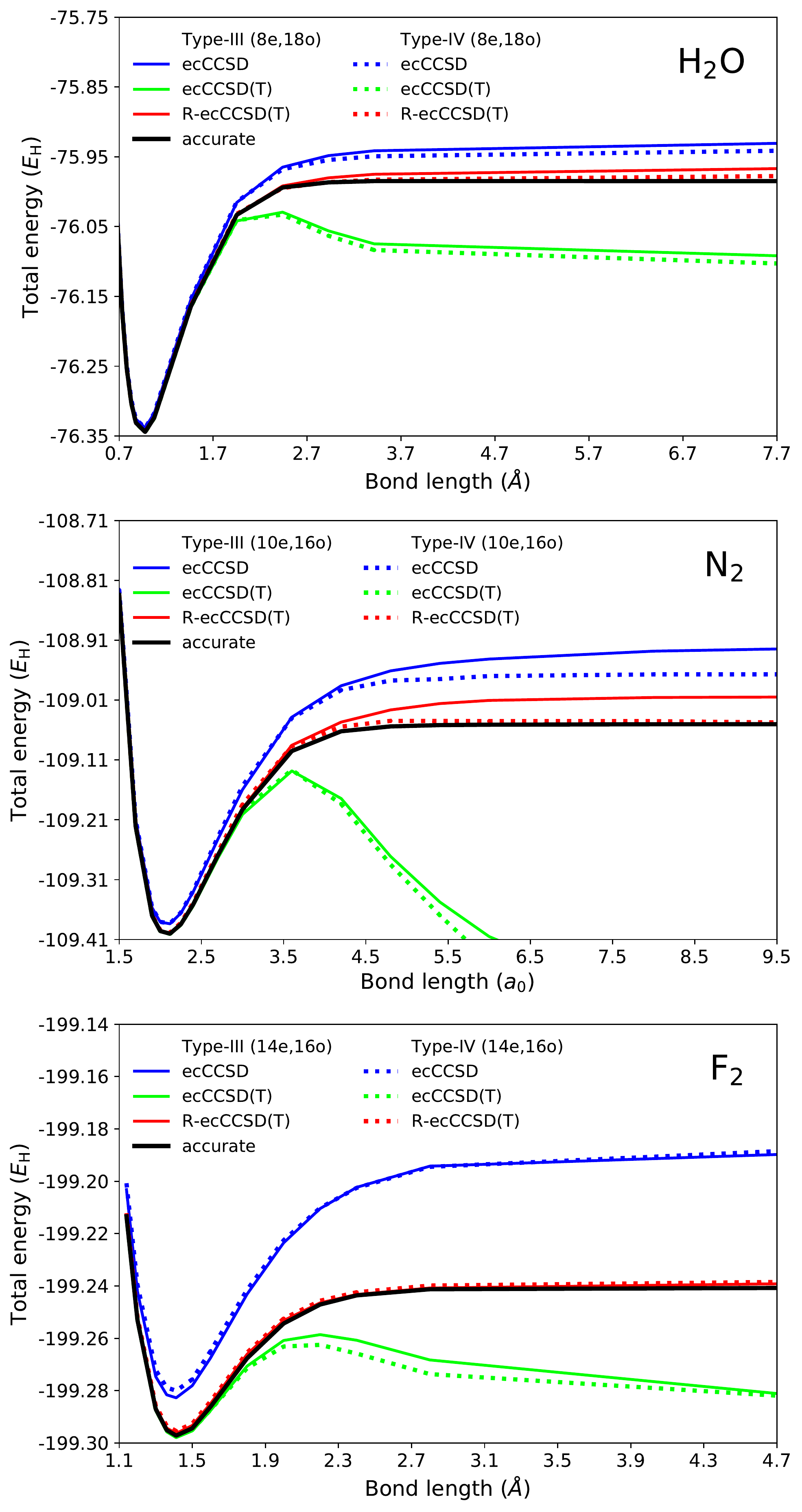} \label{fig:fpa}
\end{figure}

In addition, we investigated PESs of R-ecCCSD(T) using larger active spaces with all amplitudes (type-III) and 80\% of the amplitudes (type-IV). 
We used active spaces of (8e,18o), (10e,16o), and (14e,16o) for  H${}_2$O, N${}_2$, and F${}_2$, respectively, and obtained near exact wave functions in the spaces.
The PESs of ecCC using the type-III and type-IV external sources are shown as colored solid and dotted lines, respectively, in Figure \ref{fig:fpa}. The MAE and NPE of the PESs are given in Table \ref{table:a}.

For H${}_2$O and N${}_2$, the resulting PESs of R-ecCCSD(T) with all amplitudes (red solid lines in the top and middle panels) achieve (MAE,NPE) = ($2.9,18.0$) and ($12.1,44.3$) m$E_{\rm H}$, respectively. These are minor improvements compared to the results using the minimal active space external sources.
However, the PESs obtained using only 80\% of the external amplitudes (red dotted lines) are much closer to the black, accurate PES, and  reach a (MAE,NPE) = ($1.1,8.1$) and ($4.6,8.4$) m$E_{\rm H}$. 
{Similarly to in the minimal active space, 80\% of the amplitude weight in the larger active space corresponds to only one element of $T_4$ for H${}_2$O and the nine largest elements of $T_4$ for N${}_2$. }
When we can find such large elements, the partial usage of the amplitudes clearly has advantages in both accuracy and efficiency. It furthermore points to the importance of accounting for the error from missing relaxation in the external source amplitudes. 

Unlike in H${}_2$O and N${}_2$, there are no particularly {large} elements in the $T_3$ and $T_4$ amplitudes of F${}_2$. To truncate the amplitudes to 80\% of their weight,  we used the approximately 400 largest elements of $T_3$ and $T_4$ summing to 80\% of the total $T_3$ and $T_4$ weights at the bond length of $5.0$ \AA.
The red solid and dotted lines in the bottom panel of Figure \ref{fig:fpa} show the PESs of R-ecCCSD(T) using all and  80\% of the amplitudes, respectively.
These two PESs are very close to the accurate black curve and  reach a (MAE,NPE) = ($0.9,1.4$) and ($1.7,1.6$) m$E_{\rm H}$ for 100\% and 80\% of the amplitudes, respectively.
Although both are accurate, the partial use of the external source in this case (where there is not a small number of large elements) leads to slightly worse accuracy. 

\begin{table}[htp]
\centering
\caption{MAE and NPE (m$E_{\rm H}$) of H${}_2$O, N${}_2$, and F${}_2$ PESs in a range of geometries $R \in [0.68, 7.80]$ \AA, $R \in [1.5, 10.0]$ $a_0$, and $R \in [1.14, 5.00]$ \AA, respectively.
 ecCC methods using the near exact external sources with all elements of $T_3$ and $T_4$ amplitudes ("100\%") or only the largest elements of the $T_4$ amplitudes ("80\%") from the external source.
}
\begin{tabular}{lcccccccccc}\hline \hline
    \multirow{2}{*}{Method}  & \multirow{2}{*}{$T_3$, $T_4$} 
                                            & \multicolumn{3}{c}{H${}_2$O} & \multicolumn{3}{c}{N${}_2$} & \multicolumn{3}{c}{F${}_2$} \\ \cline{3-11} 
	                            &           &  Active & MAE   & NPE   & Active  & MAE   & NPE    & Active  & MAE  & NPE   \\ \hline
	CCSD                        &           &                           & 15.9  & 30.4  &   & 32.4  & 102.3  &   & 34.8     & 57.7 \\ \hline
    \multirow{4}{*}{ecCCSD}     & 100\%     &  \multirow{2}{*}{4e,4o}   & 24.0  & 65.9  & \multirow{2}{*}{6e,6o} & 64.1 & 149.9 & \multirow{2}{*}{2e,2o} & \multirow{2}{*}{a} & \multirow{2}{*}{a} \\
                                &  80\%     &                           & 24.0  & 65.9  &  & 57.5 & 114.5 & & & \\ \cline{2-11}
                                & 100\%     & \multirow{2}{*}{8e,18o}   & 17.6  & 49.9  & \multirow{2}{*}{10e,16o}  & 48.4  & 113.0  & \multirow{2}{*}{14e,16o}  & 25.2 & 41.4  \\
                                &  80\%     &                           & 15.6  & 39.4  &   & 41.4  & 74.0   &   & 27.0 & 40.9  \\ \hline
    R-CCSD(T)                   &           &                           & 6.8   & 43.0  &   & 32.1  & 162.1  &   &  6.2 & 10.6 \\ \hline
    \multirow{4}{*}{R-ecCCSD(T)}& 100\%     & \multirow{2}{*}{4e,4o}    & 3.9   & 23.1  & \multirow{2}{*}{6e,6o} & 12.8 & 41.4 & \multirow{2}{*}{2e,2o} & \multirow{2}{*}{a} & \multirow{2}{*}{a}\\
            	                &  80\%     &                           & 3.9   & 23.1  && 5.4  & 13.9 &&&\\ \cline{2-11}
    & 100\%     & \multirow{2}{*}{8e,18o}                               & 2.9   & 18.0  & \multirow{2}{*}{10e,16o}  & 12.1  & 44.3   & \multirow{2}{*}{14e,16o}  & 0.9  & 1.4  \\
                                &  80\%     &                           & 1.1   & 8.1   &   & 4.6   & 8.4    &   & 1.7  & 1.6  \\ \hline \hline
\end{tabular} \label{table:a}

${}^{\rm a}$ no $T_3$ and $T_4$ in the minimal active space of F${}_2$
\end{table}

\subsection{PESs with type-V external sources} \label{sec:fpal}
In the previous section, we showed that the use of larger active spaces significantly improves the ecCC PESs.
However, obtaining tightly converged DMRG wave functions in large active spaces requires more
CPU time than the subsequent CC calculation. 
For example, optimizing an external wave function with 16 orbitals requires around a few minutes of  CPU time for one DMRG sweep with $M=2000$. 
Although a few minutes is not prohibitively large in many applications, it is large compared to the subsequent ecCC calculation which only takes tens of seconds, at least for the small molecules considered here. In addition, the fact that in some cases only a partial use of the amplitudes led to better results in the last section suggests that it is not a good use of computational time to tightly converge the external source.

We thus now consider loosely converged DMRG sources (type-V) in the larger active spaces. We re-computed PESs of R-ecCCSD(T) shown in Figure \ref{fig:fpa} using a DMRG source with bond dimensions 25, 50, and 100. We used the truncation to 80\%  amplitude weight for H${}_2$O and N${}_2$, while we used all the external amplitudes for F${}_2$.
Table \ref{table:loose} shows the corresponding MAE and NPE in the same range of geometries in Table \ref{table:a}.

For all  cases, when we reduced the bond dimension to $M=100$, the MAE increased by $0.1 \sim 0.5$ m$E_H$, and the NPE increased by $0.3 \sim 1.0$ m$E_H$, compared to using $M=2000$. However for $M=100$, one DMRG sweep with 16 orbitals  took only a few seconds of CPU time at the 10$a_0$ bond length of N${}_2$, giving a better computational balance between the DMRG calculation and the subsequent ecCC calculations. When we further reduced the bond dimension to $M=25$, the MAE increased by $0.9$ m$E_H$ and the NPE increased by $0.6$ and $2.7$ m$E_H$ for H${}_2$O and N${}_2$. 
In the case of F${}_2$, which does not have a small number of large $T_3$ and $T_4$ elements, the MAE and NPE increased more, by $3.4$ and $5.4$ m$E_H$.

\subsection{PESs with type-VI external sources} \label{sec:vi}
Type VI sources use the full orbital space and employ the single HCI parameter, $\epsilon$, to select the large $T_3$ and $T_4$ amplitudes.
They do not require a choice for the CAS space, and we did not further threshold the $T_3$ and $T_4$ amplitudes.
However, for systems where the molecule has a low-spin ground state, but the dissociated
fragments have high-spin ground states, the HCI wave functions at stretched geometries
are strongly spin-contaminated when $\epsilon$ is large, even when time-reversal symmetry
is employed to reduce the spin contamination. (This type of spin-contamination is avoided in the small bond dimension DMRG calculations via the full use of spin symmetry).
Although this has little effect on the HCI energy, it makes the amplitudes unsuitable for ecCC calculations.

In the second part of Table \ref{table:loose} the MAE and NPE from type-VI sources using
$\epsilon= 0.01$ and $0.003$ are shown.  These are evaluated over a smaller range of
geometries than those used in the first part of Table \ref{table:loose} for the reason mentioned
above.
Since N$_2$ is a singlet that dissociates into atoms that are quartets, it has particularly
large errors.  The errors improve upon going from $\epsilon=0.01$ to $\epsilon=0.003$, in particular
the PEC has an unphysical dip at $\epsilon=0.01$ which disappears at $\epsilon=0.003$, but
the errors are still substantial.
Similarly to how using all the amplitudes could lead to larger errors than only using some of the amplitudes in the previous sections, the errors incurred from the type VI sources here emphasize that the  quality of the external amplitudes cannot be judged solely from the energy obtained by the external method. 

\begin{table}[htp]
\centering
\caption{MAE and NPE (m$E_{\rm H}$) of R-ecCCSD(T) PESs for H${}_2$O, N${}_2$, and F${}_2$ using the approximate DMRG and HCI sources with a small bond dimension ($M$) or large threshold ($\epsilon$). 
Detailed explanation to be found in the main text.}
\begin{tabular}{lccccccccc}\hline \hline
    \multicolumn{1}{c}{\multirow{2}{*}{$M$}} & \multicolumn{3}{c}{H${}_2$O} & \multicolumn{3}{c}{N${}_2$} & \multicolumn{3}{c}{F${}_2$} \\ \cline{2-10} 
	     &  Active & MAE ${}^a$ & NPE ${}^a$ &  Active  & MAE ${}^a$ & NPE ${}^a$  &  Active  & MAE ${}^a$ & NPE ${}^a$ \\ \hline
    25   &  \multirow{4}{*}{(8e,18o)} & 3.8 & 8.7 & \multirow{4}{*}{(10e,16o)} & 5.5 & 11.1 & \multirow{4}{*}{(14e,16o)} & 4.3 & 6.8 \\
    50   &         & 3.4 & 8.2 &          & 5.2 & 9.6  &          & 3.5 & 3.4 \\
    100  &         & 3.0 & 8.4 &          & 4.9 & 9.4  &          & 1.4 & 2.1 \\
    2000 &         & 2.9 & 8.1 &          & 4.6 & 8.4  &          & 0.9 & 1.4 \\ \hline
\multicolumn{1}{c}{\multirow{2}{*}{$\epsilon$}} & \multicolumn{3}{c}{H${}_2$O} & \multicolumn{3}{c}{N${}_2$} & \multicolumn{3}{c}{F${}_2$} \\ \cline{2-10}
         &  Active & MAE ${}^b$ & NPE ${}^b$ &  Active    & MAE ${}^b$ & NPE ${}^b$ &  Active   & MAE ${}^b$ & NPE ${}^b$ \\ \hline    
$0.01$   &  \multirow{2}{*}{(10e,58o)} & 1.6 & 7.1 & \multirow{2}{*}{(14e,60o)} & 8.1 & 61.5 & \multirow{2}{*}{(14e,58o)} 
                                                                    & 3.6 & 4.7 \\  
$0.003$  &         & 1.3 & 4.4 &            & 3.6 & 15.8 &          & 3.0 & 5.8 \\ \hline \hline
\end{tabular} \label{table:loose}

${}^a$ $R \in [0.68, 7.80]$ \AA, $[1.5, 10.0]$ $a_0$, and $[1.14, 5.00]$ \AA $\ $ for H${}_2$O, N${}_2$, and F${}_2$ PESs, respectively.  \\
${}^b$ $R \in [0.68, 3.41]$ \AA, $[1.5, 5.4]$ $a_0$, and $[1.14, 2.80]$ \AA $\ $ for H${}_2$O, N${}_2$, and F${}_2$ PESs, respectively.  
\end{table}

\subsection{Error analysis} \label{sec:diag}

\begin{figure}[htp]
\centering
\caption{Absolute errors of R-CCSD(T) and ecCC methods on a log scale plotted against the $D_2$ diagnostic. Detailed explanation in the  main text. }
\includegraphics[width=\textwidth,height=\textheight,keepaspectratio]{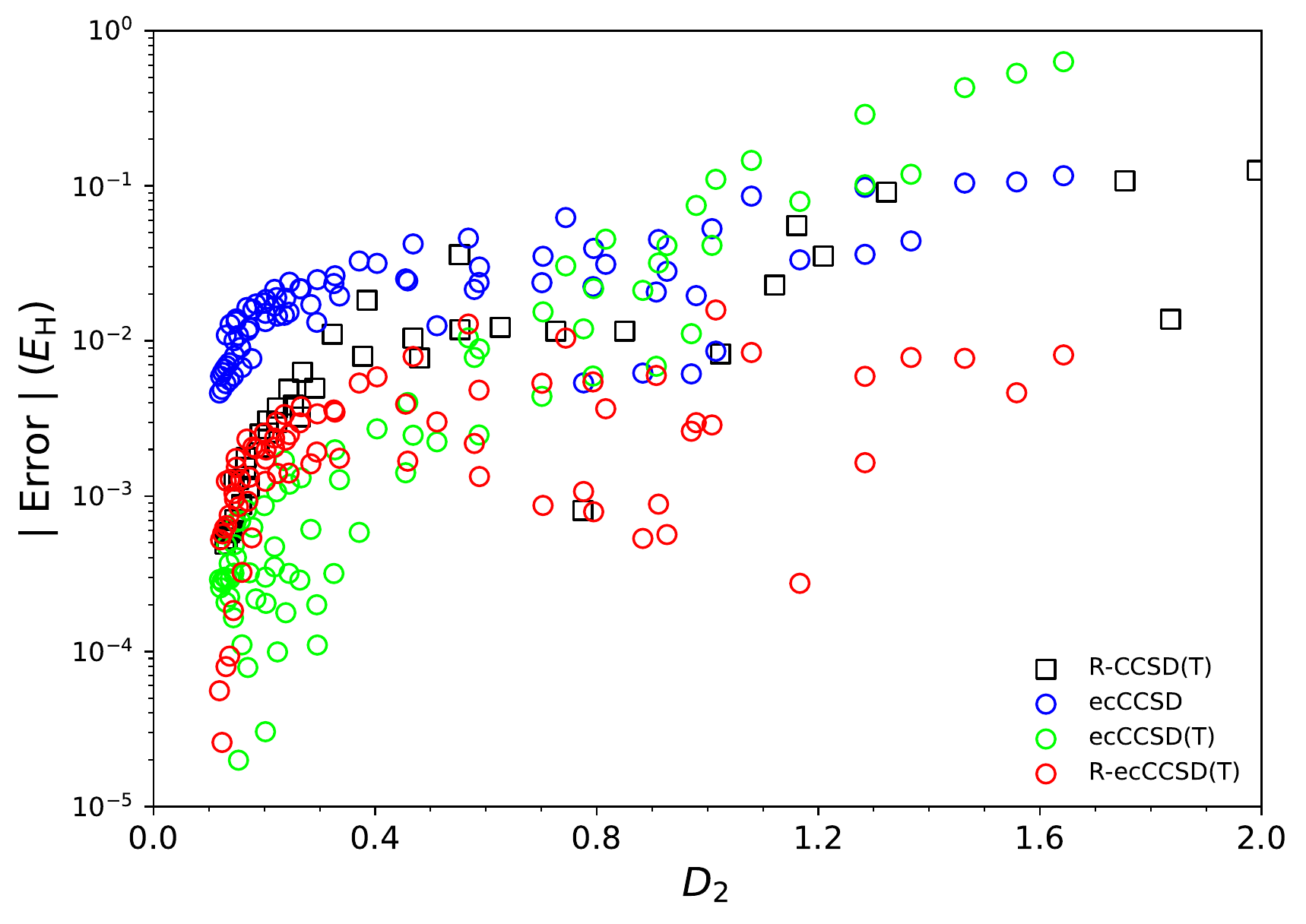} \label{fig:diag}
\end{figure}

In this section, we present the errors of R-ecCCSD(T) for the systems in this work and analyze them using the well known CC error diagnostic $D_2$, defined by the matrix 2-norm of the $T_2$ amplitudes.~{\cite{nielsen1999double}} The magnitude of $D_2$ can be used to distinguish between the SR and MR character of the different geometries on the PESs.
Organic molecules are sometimes considered to have MR character when $D_2$ is larger than $0.18$.~{\cite{nielsen1999double}}
Figure \ref{fig:diag} shows the absolute errors on a log scale versus the $D_2$ diagnostic.
Each symbol represents a geometry on the PESs of one of the molecules, H${}_2$O, N${}_2$, and F${}_2$.
Square symbols denote R-CCSD(T) and circle symbols denote the various externally corrected theories, using the type-V and the type-VI external sources.
For $D_2$ ranging from $0$ to $1.7$, the absolute errors of R-ecCCSD(T) (red circles) are less than $0.015$ E${}_H$ and most of the errors are smaller than those of R-CCSD(T) (black squares) and ecCCSD (blue circles).
The absolute errors of R-ecCCSD(T) are less than those of ecCCSD(T) (green  circles) in the MR region where $D_2 > 0.4$, while they are {mostly} greater than those of ecCCSD(T) in the range $0.0 < D_2 < 0.4$.
Overall, it is clear to see that R-ecCCSD(T) offers the most balanced treatment of errors across a wide range of SR and MR character. However, for very weakly correlated systems, the original (T) correction is slightly more accurate than the renormalized (T) correction.

\subsection{Limitations} \label{sec:limit}

Although the work above shows that it is possible to obtain quantitative accuracy across the full potential energy surface, at a reasonable computational cost, using the combination of external sources and the renormalized (T) correction,  SR ecCC approaches have a fundamental limitation when the CI coefficient of the reference configuration (the HF configuration in this work) in the external source is exactly or numerically zero.
This leads to overflow due to the assumption of intermediate normalization. 
The smallest reference coefficient value we encountered in this work was $0.2$ at the stretched  $10a_0$ geometry of N${}_2$. 
{Although this coefficient is not numerically zero,}
it is difficult to converge the ecCCSD energy. We extracted initial amplitudes of $T_1$ and $T_2$ from the external source and then iteratively converged the ecCCSD energy using a damping parameter of $0.05$ to update the amplitudes. (We did not use  the direct inversion in the iterative subspace (DIIS) algorithm). 
At this geometry, the energy could be converged monotonically up to a threshold of $10^{-5}$ Hartrees, although the norm of the amplitudes could not be converged, and we simply used amplitudes from the last iteration in the set of monotonically decreasing energies. 

\section{Conclusion} \label{sec:conclusions}
In this work, we explored externally corrected coupled cluster with a renormalized triples correction (R-ecCCSD(T))  using DMRG and HCI and external sources. The critical question  is how to best balance the accuracy and cost of computing the external source with the cost of the overall method. To this end, we  considered multiple types of external sources: "exact" external sources, where the DMRG and HCI wavefunctions were tightly converged, and "approximate" external sources where they were loosely converged.
We also considered both "full" usage of the $T_3$ and $T_4$ amplitudes,
and "partial" usage, wherein we retained only the largest elements. 

For all systems considered here, we found that R-ecCCSD(T) can significantly improve on the description of the potential energy surface given either by the external source alone or CC alone. For example, the unphysical dips in the PES of H${}_2$O and N${}_2$ in the bond-stretched region can be completely eliminated.
The use of approximate external sources, possibly with truncation to only the {large} $T_3$ and $T_4$ amplitudes, appears to be a practical way to balance the cost of the external calculation and the coupled cluster calculation in small molecules.
Using the $D_2$ diagnostic to characterize the different points on the potential energy surfaces, we find that R-ecCCSD(T) gives absolute errors of less than 15 m$E_H$ in the range of $D_2$ from $0.0$ to $1.7$. In fact, the errors of R-ecCCSD(T) are less than those of ecCCSD and R-CCSD(T) in almost all cases, except when $D_2$ is very small, where the renormalized (T) correction appears to be slightly less accurate than the simple (T) correction. 

There are several interesting questions remaining which lie beyond what we have considered in this work. For example, while R-ecCCSD(T) appears quite stable up to large values of the $D_2$ diagnostic, what is the largest amount of multi-reference character which can be handled? Here, the difficulty in solving the CC equations, and the divergence of the amplitudes reflecting the problems of intermediate normalization, cannot be ignored. In addition, in the realm of quasidegenerate problems, we can ask whether other non-iterative corrections such as the "completely renormalized" triples and quadruples corrections,~\cite{kowalski2000mmcc} corresponding to CR-ecCCSD(T) and CR-ecCCSD(T,Q), would further improve on the present R-ecCCSD(T) method. Finally, the current approximation, with its modest computational requirements on the external source, is applicable to the same scale of systems that can be handled by single reference coupled cluster methods. Thus understanding the performance of this method in larger correlated systems is of interest. 

\textit{Note}: As this work was being prepared for submission, we were made aware of a related recent submission to the arxiv~\cite{magoulas2021externally}, that also discusses  selected configuration interaction as an external source and perturbative triples corrections in the context of externally corrected coupled cluster methods. 

\begin{acknowledgement}
Work by SL, HZ and GKC was supported by the US National Science Foundation via Award CHE-1655333. GKC is a Simons Investigator.
SS was supported by the US National Science Foundation grant CHE-1800584 and the Sloan research fellowship.
CJU was supported in part by the AFOSR under grant FA9550-18-1-0095.
\end{acknowledgement}

\begin{suppinfo}

\end{suppinfo}

\clearpage
\bibliography{refs}

\clearpage
\appendix
\addcontentsline{toc}{section}{Appendices}
\section*{Appendix}

\section{A sweep algorithm converting MPS to CI coefficients with a threshold} \label{apdx:mpsci}

\setcounter{equation}{0}
\renewcommand{\theequation}{A.\arabic{equation}}

In DMRG, the electronic wave function is represented by a matrix-product state (MPS)
\begin{align}
    | \Psi \rangle & = \sum_{n_1, \dots, n_K} \sum_{\alpha_1, \dots, \alpha_{K-1}} A^{n_1}_{\alpha_1} A^{n_2}_{\alpha_1 \alpha_2} \cdots A^{n_K}_{\alpha_{K-1}} | n_1 n_2 \cdots n_K \rangle, \\
    \{ n \} & = \{ {\rm vac} ,  \uparrow, \downarrow, \uparrow \downarrow \},
\end{align}
where $n_i$ is the occupation of orbital $i$, $| n_1 n_2 \cdots n_K \rangle$ is the occupation-number representation of a determinant, and $\alpha_i$ is an auxiliary index. Here, $\sum_{\alpha_i} A^{n_i}_{\alpha_{i-1}\alpha_i} A^{n_{i+1}}_{\alpha_{i}\alpha_{i+1}}$ denotes a matrix product and it is assumed that the dimensions of all the auxiliary indices are the same (the bond dimension $M$).
The wave function of the ground state can be optimized by the efficient DMRG sweep algorithm. \cite{}

In order to get $Q^{\rm ec}_3 T_3$ and $Q^{\rm ec}_4 T_4$ in Eq. (\ref{eq:t}), we extract quadruple- and lower-order CI excitation amplitudes  from the MPS by a sweep algorithm. To avoid repeated or unnecessary computation, we here describe how to obtain  CI coefficients whose values are larger than a threshold in Eq. (\ref{eq:dmrg_V}), with a concomitant reduction in computational cost and memory usage from a naive approach. 

We first start with the MPS in left canonical form. During a sweep to compute the amplitude, at any given point (e.g. at some site $p$) one has a set of partial coefficients $c_{\alpha_p}(n_1 n_2 \ldots n_p) = \sum_{\alpha_1 \ldots \alpha_{p-1}} A^{n_1}_{\alpha_1} \ldots A^{n_p}_{\alpha_{p-1}\alpha_p}$. If $\sum_{\alpha_{p}}|c_{\alpha_p}(n_1 n_2 \ldots n_p)|^2 < \text{thresh}$, then this partial coefficient is dropped, as are all determinants involving the occupancy string $|n_1 n_2 \ldots n_p\rangle$. This is because if the MPS is in left canonical form, the above condition on the partial coefficient guarantees that the coefficient of any  determinant generated by the MPS which contains $|n_1 n_2 \ldots n_p\rangle$ as a substring is also less than the threshold in magnitude. In addition, since the orbitals are associated with definite hole or particle character, we also drop any coefficient associated with more than 4 holes or 4 particles. Finally, in this process, we can take advantage of the conserved quantum numbers to only generate symmetry unique partial coefficients (e.g. if $S_z=0$, then the values of $c_{\alpha_{p+1}}$ come in time-reversal pairs and only one needs to be considered). 
Thus using the above algorithm we can completely avoid generating any determinants with coefficients below the threshold.

\end{document}


\begin{table}[htp]
\centering
\caption{Reference SHCI energies of H${}_2$O and N${}_2$ in the full space of the cc-pVTZ basis set, without frozen orbitals. 
The SHCI calculations used $\epsilon_1=4\times 10^{-5}$ for the variational calculation and $\epsilon_2=10^{-8}$ for the stochastic perturbative calculation. The statistical uncertainty is represented as a number in parentheses. 
The error in the energies is significantly smaller than the deviation from any of the methods discussed in this work.} 
\begin{tabular}{clcl}\hline \hline
\multicolumn{2}{c}{H${}_2$O} & \multicolumn{2}{c}{N${}_2$} \\
$R / R_e$ ${}^a$  & \multicolumn{1}{c}{$E$ (Hartree)} & $R / a_0$ & \multicolumn{1}{c}{$E$ (Hartree)} \\ \hline
0.70 &  -76.048012(3)  &   1.50  & -108.834070(4) \\
0.75 &  -76.171478(3)  &   1.70  & -109.220632(4) \\
0.80 &  -76.252531(3)  &   1.90  & -109.370465(6) \\
0.85 &  -76.302894(5)  &   2.00  & -109.39571(3)  \\
0.90 &  -76.331176(3)  &   2.12  & -109.400421(7) \\
1.00 &  -76.344206(2)  &   2.25  & -109.38454(6)  \\
1.10 &  -76.323868(3)  &   2.40  & -109.35158(1)  \\
1.50 &  -76.164352(3)  &   2.70  & -109.26945(1)  \\
2.00 &  -76.033235(2)  &   3.00  & -109.19275(1)  \\
2.50 &  -75.994010(2)  &   3.60  & -109.09515(2)  \\
3.00 &  -75.986708(1)  &   4.20  & -109.06218(2)  \\
3.50 &  -75.9848292(9) &   4.80  & -109.05417(2)  \\
8.00 &  -75.9848822(3) &   5.40  & -109.05183(2)  \\
     &                 &   6.00  & -109.05099(2)  \\
     &                 &   8.00  & -109.05028(1)  \\
     &                 &   10.00 & -109.05020(1)  \\ \hline \hline
\end{tabular}

${}^a$ The equilibrium bond length $R_e$ of H${}_2$O is $0.97551$\AA.
\end{table}